\documentstyle[11pt]{article}
\input epsf
\def\sdtimes{\mathbin{\hbox{\hskip2pt\vrule
height 4.8pt depth -.3pt width .20pt\hskip-2pt$\times$}}}
\begin{document}
\thispagestyle{empty}
\baselineskip 20pt
\rightline{CU-TP-842}
\rightline{hep-th/9706023}
\
\vskip 1cm
\centerline{\Large\bf A Family of $N=2$ Gauge Theories } 
\centerline{\Large\bf with Exact S-Duality}
\vskip  1cm
\centerline{Kimyeong Lee\footnote{electronic mail: klee@phys.columbia.edu} 
and Piljin Yi\footnote{electronic mail: piljin@phys.columbia.edu}}
\vskip 2mm
\centerline{\it Physics Department, Columbia University, New York,  NY 10027}
\vskip 2cm
 
\centerline{\bf ABSTRACT}
\vskip 1cm
\begin{quote}
We study an infinite family of $N=2$ $Sp(2n)$ gauge theories that
naturally arise from the D3-brane probe dynamics in F-theory.  The
matter sector consists of four fundamental and one antisymmetric
tensor hyper multiplets. We propose that, in the limit of vanishing
bare masses, the theory has  exact $SO(8)\sdtimes SL(2,Z)$
duality. We examine the semiclassical BPS spectrum in the Coulomb
phase by quantizing various monopole moduli space dynamics, and show
that it is indeed consistent with the exact S-duality.

\end{quote}

\newpage
\section{Introduction}

Electromagnetic duality is a powerful idea that relates strongly
coupled theories to weakly coupled ones.  Montonen and Olive's
conjecture of the duality has been shown to be realized very simply in
the $N=4$ supersymmetric Yang-Mills theory~\cite{montonen}. This
theory is finite and so the duality is expected to be exact in all
energy scales. Sen was the first to realize that the larger $SL(2,Z)$
duality holds in this theory with gauge group $SU(2)$~\cite{sen}. Such
S-duality has been also studied with higher rank groups. Furthermore,
Seiberg and Witten proposed that a $N=2$ supersymmetric Yang-Mills
theory with $SU(2)$ gauge group and four massless flavors of matter
fields in the fundamental representation, also has the exact S-duality
in the Coulomb phase~\cite{seiberg}. This conjecture was further
supported from the study of the low energy quantum mechanics of dyons
of magnetic charge one and two~\cite{gaunt1} and from the study of
effective action~\cite{ferrari}.

Since then, there have been attempts to find to other $N=2$ models
beyond $SU(2)$ gauge group (such as $SU(3)$) with an exact S-duality,
none of which succeeded \cite{cederwall,aharony}.  It has been also
suggested that maybe the S-duality of the $SU(2)$ case is  accidental
and the scale-invariant $N=2$ models need not be S-dual in general.
In this work we study an infinite family of scale-invariant $N=2$
supersymmetric Yang-Mills theory, and propose that they all possess
exact S-duality in much the same way as the scale-invariant
Seiberg-Witten $SU(2)$ does. The proposed theory has the gauge group
$Sp(2n)$ with a single antisymmetric tensor and four fundamental
matter fields.  We will provide substantial evidences supporting the
exact S-duality by studying the multi-monopole dynamics and the
resulting semiclassical spectrum. Seiberg and Witten's $SU(2)=SP(2)$
model is clearly the first of the series of our model since the
antisymmetric tensor of $Sp(2n)$ is a singlet and decouples for $n=1$.

This particular generalization is in fact well-motivated by
considering the simplest compactification of F-theory
\cite{Ftheory,sen2,banks,sen3}.  In the simplest reincarnation of
F-theory, one is effectively compactifying a type IIB theory on a two
sphere but in such a manner that the dilaton-axion field varies with the
holomorphic coordinate on the sphere. The consistency requires 24
singular points on the sphere, which extend to the remaining 7 spatial
dimensions and can be thought of as a kind of D7-branes. Sen observed
\cite{sen2} that, when the sphere gets large and 24 singular points
are split into four groups of 6 singularities, each congregating
together far away from the others, the local geometry of the two
sphere near each group is identical to the vacuum moduli space of a
$N=2$ $SU(2)$ model with four (massive) fundamental hyper multiplets,
the exact solution of which was found by Seiberg and Witten some time
before. This remarkable fact was quickly explained by the observation
that the $SU(2)$ theory in question is simply a world-volume
Yang-Mills theory on a probe D3-brane parallel to D7-branes~\cite{banks}.

The limit of massless hyper multiplets is achieved by letting each
group of the six nearby singularities to coincide at four different
points. In this limit, the string background being probed by the
D3-brane is rather trivial. There is no Ramond-Ramond charge that may
transform nontrivially under the $SL(2,Z)$ U-duality of the type IIB
string, and the dilaton-axion becomes uniform.  As a matter of fact,
the theory is an orientifold of type IIB theory with all 16 D7-branes
localized at the four orientifold planes, and is dual to a good old
orbifold $T^2/Z_2$ compactification of the type I string. Since this
string background is invariant under the $SL(2,Z)$ U-duality of the
type IIB string up to a trivial redefinition of the coupling, one
could argue that the D3-probe dynamics must be similarly invariant
under certain S-duality.

Recently, by generalizing the above idea to include the $n$ D3-brane
probes, two independent groups studied a series of $N=2$ world-volume
theory on the D3-branes~\cite{spn}.  The gauge group is $Sp(2n)$, and
the matter multiplet includes four fundamental and a single
antisymmetric tensor. Again, in the limit where quartets of D7-branes
coincide with the orientifold plane, one would expect to recover a
theory with an exact S-duality. This provides an excellent motivation
for studying the proposed generalization of the scale-invariant
Seiberg-Witten model.

This F-theory consideration suggests the whole family of $Sp(2n)$
models share the same duality group. With $n=1$, the duality group is
known to be $SO(8)\sdtimes SL(2,Z)$. The $SO(8)$ is the global
symmetry of the theory for all $n$, and rotates the fundamental hyper
multiplets and their charge conjugates. This $SO(8)$ acts on neither
the adjoint nor the antisymmetric fields. As in the $n=1$ case
previously studied, we will find that the $SL(2,Z)$ action involve a
permutations of the vector $8_v$, the spinor $8_s$ and the chiral
$8_c$ representations of $SO(8)$. When the fundamental matters
disappear for example by becoming infinitely massive, the duality
group, which now has a new interpretation as a monodromy group,
collapses to $\Gamma(2)$. The nontrivial triality action of $SL(2,Z)$
is naturally encoded in the fact that the six coset elements
$SL(2,Z)/\Gamma(2)$ form a permutation group of $(8_v,8_s,8_c)$.

Our goal here is to confirm that the semiclassical BPS spectrum is
indeed invariant under such S-duality, by quantizing various
multi-monopole dynamics. One complication of having a larger gauge
group is that the Coulomb phase structure is rather complicated even
after assuming the maximal breaking to $U(1)^n$. There are two adjoint
Higgs fields in $N=2$ vector multiplet, and their vacuum expectation
values must commute with each other. But when the gauge group has rank
larger than one, the two Higgs expectations need not be parallel
within the Cartan subalgebra \cite{aharony,hollowood}. Depending on
whether they are aligned or not, the multi-monopole dynamics will
differ drastically. The 4 new Higgs from
the antisymmetric hyper multiplets can make matters even more complicated.

When two vacuum values of the Higgs fields are not parallel, a
monopole corresponding to each and every root is fundamental in the
sense that there are only four bosonic zero modes. In order to test
the self-duality of the BPS spectra, it is sufficient to consider the
interactions of identical monopoles. By the fermion zero mode analysis
we show that in this case the degeneracy and the dyonic excitations of
a single monopole match exactly the prediction of the duality.  In
fact, more generally, the problem of counting states can be mapped
either to the $N=2$ $SU(2)$ case studied by Gauntlett and Harvey as
well as Sethi, Stern and Zaslow~\cite{gaunt1} or (very surprisingly)
to a $N=4$ $SU(2)$ case studied by Sen and subsequently by
others~\cite{sen,segal}.  In this way, we will find that the S-duality
of the scale-invariant Seiberg-Witten $SU(2)$ model combined with the
S-duality of the $N=4$ $SU(2)$ guarantees that the same is true for
the $N=2$ $Sp(2n)$ theories for all $n$. As will be clear later on,
such a reduction of the monopole dynamics to other known problems is
the prevailing theme throughout this work.

In such generic vacua, the problem of finding BPS spectrum can be carried
out from the F-theory viewpoint as in Ref.~\cite{sen3}. However, this 
method is ineffective in probing threshold bound states that necessarily
arise when the vacuum expectation values are aligned; Because $N=2$ 
theories are known to have discontinuities in the dyon spectrum, one must 
check for the dyon spectrum explicitly in all cases~\cite{seiberg}.

When the two expectation values are parallel, the magnetic monopole moduli
space becomes quite rich. We can choose $n$ fundamental monopoles,
from which a configuration of any magnetic charge can be built 
\cite{weinberg}.  The bosonic moduli space of arbitrary number of distinct 
monopoles has been found \cite{klee1,murray}. 
Studying the index bundles on such moduli space, we can
again relate many of the bound state problems to those found in $N=4$
Yang-Mills theory \cite{gaunt2,gibbons}. 
Similarly as above, at least a substantial part  
of the S-dual BPS spectrum can be shown to arise from the seemingly
unrelated fact that $N=4$ Yang-Mills theory has the exact S-duality of
$SL(2,Z)$.

The plan of this work is as follows. In Section 2, we briefly review
properties of the $SP(2n)$ gauge algebra and the elementary matter content. 
For simplicity, we shall set the antisymmetric Higgs expectation to
zero without a loss of generality. In Section 3, we consider the magnetic 
monopole spectrum when the vacuum expectation values of two Higgs fields 
are not aligned. In Section 4, we 
consider special cases when the expectation values are 
parallel. The moduli space dynamics are far more subtle here, but
turned out be to tractable in many cases. We review some basic fact about
the index bundle spanned by the fermionic zero modes, which will help us
to relate the present bound state problems to well-known ones in some
$N=4$ theories. We conclude in Section 5 with some remarks.

\section{The $Sp(2n)$ Lie Algebra}

In this section, we  set up the notation and describe the matter 
content briefly. The roots and the weights of $Sp(2n)$ algebra are easily
described by $n$ orthonormal vectors $e_i$ in $R^n$ such that $e_i\cdot e_j
=\delta_{ij}$. With the Cartan generators $H_i$, $i=1,\dots,n$ 
normalized by
\begin{equation}
{\rm tr} \,H_iH_j=\delta_{ij},
\end{equation}
the $2n^2$ roots of $Sp(2n)$, or equivalently the weights of the adjoint 
representation, are given by $\{\alpha\}=\{\gamma_i,\beta_{ij}^\pm,
-\gamma_i,-\beta_{ij}^\pm\}$ where we have defined
\begin{eqnarray}
\gamma_i&=& \sqrt{2} \,e_i, \\
\beta_{ij}^\pm&=&\frac{1}{\sqrt{2}}\,(e_i\pm e_j)\quad i>j.
\end{eqnarray}
Furthermore, the subset $\{ \beta_{ij}^\pm,-\beta_{ij}^\pm\}$ 
contains all $2n^2-2n$ nonzero weights of an antisymmetric tensor, while the
$2n$-dimensional fundamental representation has the weights $\{\gamma_i/2,
-\gamma_i/2\}$. For all three representations, the multiplicity 
is one for any nonzero weight. 

\begin{center}
\leavevmode
\epsfysize=4in
\epsfbox{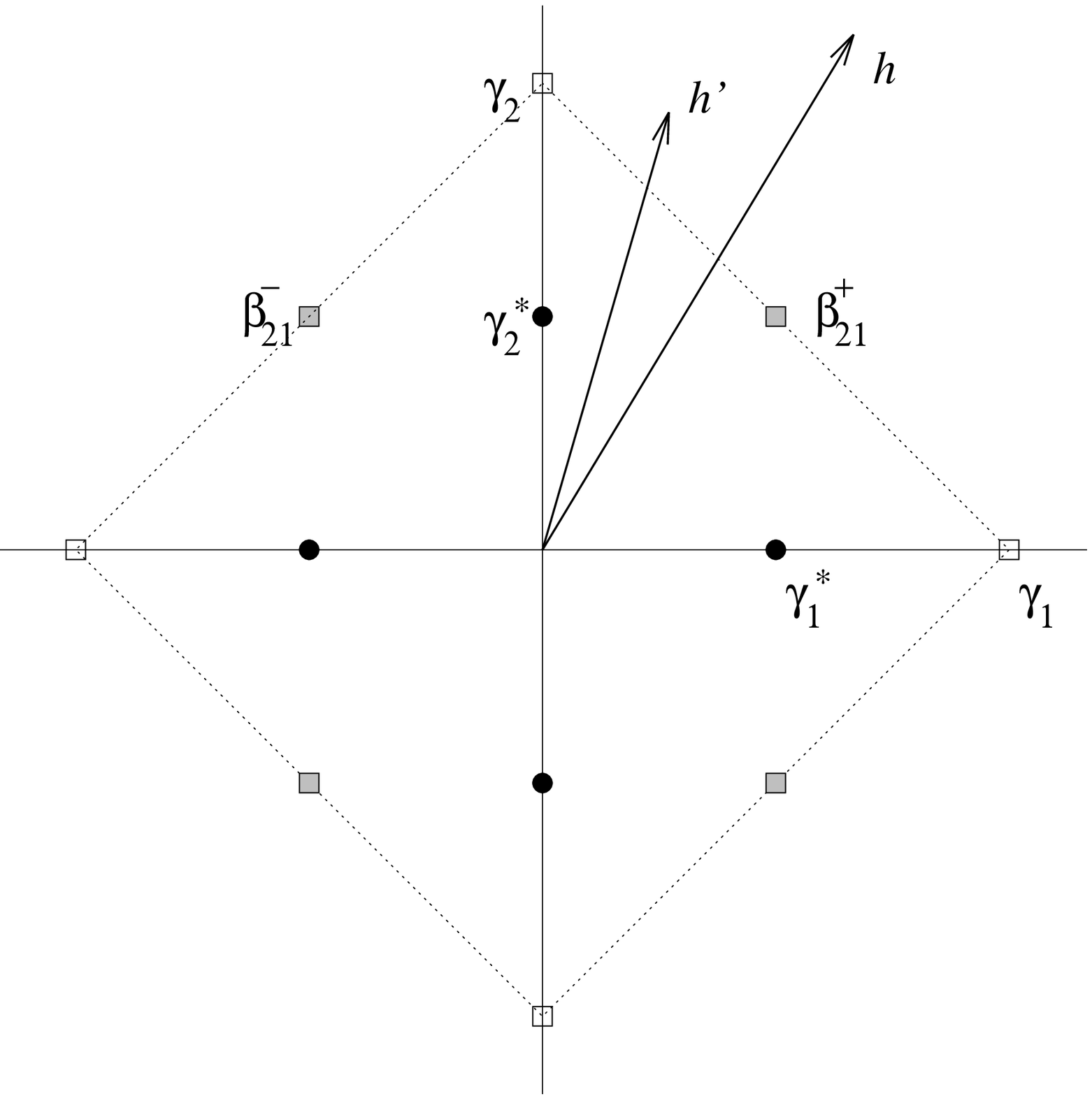}
\end{center}

\begin{quote}
{\bf Figure 1:} A diagram that shows the relevant weights of $Sp(4)$ 
gauge algebra. The eight roots are found along the dotted line.
\end{quote}

The gauge symmetry is spontaneously broken to $U(1)^n$ for adjoint
generic Higgs values, and in the unitary gauge where the $H_i$'s are
all unbroken the particles associated with nonzero weights acquire a
charge and a mass.  The elementary spectrum consists of an
antisymmetric tensor and four fundamental fields in a hyper multiplet
as well as adjoint gauge particles in a vector multiplet.  All
particles of zero weight are chargeless and massless, and so belongs
to a short supermultiplet. These neutral particles are invariant under
the $SL(2,Z)$ transformation.  We are interested in (charged) BPS
spectra, which transforms under the duality, so it is useful to
delineate the matter content for each nonzero weight.  Breaking up
these elementary $Sp(2n)$ multiplets with respects to the unbroken
$U(1)^n$, we find the following distributions of $U(1)^n$ charged
$N=2$ supermultiplets,
\begin{eqnarray}
\pm\gamma_i/2&\rightarrow & \hbox{4 hypers}  \nonumber\\ 
\pm\gamma_i&\rightarrow & \hbox{1 vector} \nonumber \\
\pm\beta_{ij}^- &\rightarrow & \hbox{1 vector + 1 hyper} \nonumber \\
\pm\beta_{ij}^+ &\rightarrow & \hbox{1 vector + 1 hyper} \label{particle}
\end{eqnarray}
Note that each of these charged elementary particles saturate the BPS
bound, and thus preserve half of the supersymmetry. The theory
possesses the famous $SO(8)$ global symmetry that acts on the
fundamental hyper multiplets. Because the fundamental representation
is pseudo-real, we can effectively consider the 4 hyper multiplets as
being composed of 8 half-hyper multiplets, which transform in the
vector representation $8_v$ of the $SO(8)$. In the following sections,
we will show that this elementary spectrum is in fact a part of the
full S-dual BPS spectrum.

\section{Self-dual BPS Spectra in Generic Vacua}

\subsection{Classical Solitons and Fermionic Zero Modes}

Let the two Higgs expectations in the unitary gauge 
be denoted by $h_i$ and $h_i'$. The BPS mass
formula for a purely magnetic soliton of charge $\alpha^*=\alpha/
\alpha^2$ is 
\begin{equation}
M_{\alpha^*}=4\pi\,\sqrt{(h\cdot\alpha^*)^2+(h'\cdot\alpha^*)^2}.
\end{equation}
Suppose one can write $\alpha^*=\mu^*+\nu^*$ for some roots $\mu$ and $\nu$.
For generic $h$ and $h'$, the mass formula is not additive and one
has \cite{aharony,hollowood}
\begin{equation}
M_{\alpha^*}\le M_{\mu^*}+ M_{\nu^*},
\end{equation}
where the equality cannot be saturated if
\begin{equation}
(h\cdot\mu^*)(h'\cdot\nu^*)\neq (h'\cdot\mu^*)(h\cdot\nu^*) .
\end{equation}
The condition $(h\cdot\mu^*)(h'\cdot\nu^*)= (h'\cdot\mu^*)(h\cdot\nu^*)$ is 
satisfied only on a measure zero subset of the vacuum moduli space where
$h$ and $k$ are parallel in the $(\mu,\nu)$ plane, so
in a generic vacuum the $\alpha^*$ monopole is classically stable against 
a decay into a $\mu^*$ monopole and a $\nu^*$ monopole.

Even in such a generic vacuum, it is still possible to embed the spherically
symmetric $SU(2)$ BPS monopole solution to a larger gauge group, along any 
given root $\alpha$. Using the $U(1)$ $R$-symmetry of the $N=2$ superalgebra,
one can rotate the Higgs expectation value $h+ih'$ so that $
(h\cdot\alpha^*)+i(h'\cdot\alpha^*)$ is real. Once this is achieved, the
embedding of the $SU(2)$ soliton may proceed as if there is only one 
Higgs field. The embedded subgroup generated by 
\begin{eqnarray} 
t^1(\alpha) &=& \frac{1 }{ \sqrt{2{\alpha^2}}} 
        \left(E_{\alpha} + 
        E_{-\alpha} \right), \nonumber \\
t^2(\alpha) &=& -\frac{i}{ \sqrt{2{\alpha^2}}} 
        \left(E_{\alpha} - 
        E_{-\alpha} \right), \nonumber \\
t^3(\alpha) &=&  {\alpha^*}  \cdot  { H} . 
\label{sub}
\end{eqnarray}
will be denote by $SU(2)_\alpha$ in this note. Denoting the $SU(2)$
solution with one Higgs by $A_i^s(r;v)$ and $\Phi^s(r;v)$ with the symmetry
breaking scale $v$, the embedded solution for the root $\alpha$ is
\begin{eqnarray}
A_i(r) &=& A_i^s(r;h\cdot\alpha)t^s(\alpha), \nonumber \\
\Phi(r) &=&
\Phi^s(r;h\cdot\alpha)t^s(\alpha)+(h-h\cdot\alpha^*\,\alpha)\cdot H.
\end{eqnarray}
The other Higgs field $\Phi'$ is frozen at $h'\cdot H$ once we insure
$h'\cdot\alpha=0$. {}From this, we can easily see that the magnetic charge 
of such a soliton is $\alpha^*$, in accordance with the Dirac quantization
condition. The  bosonic zero modes around 
such a soliton must only consist of the 3 translational degrees of freedom 
and a single $U(1)$ phase, and by the supersymmetry it follows that there 
are only 2 fermionic zero modes from the Dirac spinor field in the adjoint
vector multiplet.

The zero mode analysis of BPS monopoles with one real adjoint Higgs was 
carried out by E. Weinberg. Adding an extra Higgs field is especially 
simple in the background of an $SU(2)_\alpha$ embedded monopole. 
As above, let $\Phi$ and $\Phi'$ be the two Higgs whose expectation values 
are  $h$ and $h'$, respectively, and we use the global $R$-symmetry to
attain $h'\cdot\alpha=0$. For the $SU(2)_\alpha$ embedded monopoles, 
the relevant zero mode equations can be written in the spinorial 
form as \cite{weinberg,hollowood},
\begin{equation}
[\gamma^i D^i+\gamma^5 \Phi_\alpha]\, \Psi +
\{h'\cdot H+\gamma^5(h\cdot\alpha)\,y\}\,\Psi=0,
\end{equation}
where the operator inside the bracket is built from the $SU(2)_\alpha$
part of the embedded monopole, and the hypercharge $y$ reflects the 
discrepancy between the Higgs expectation $h$ and its $SU(2)_\alpha$ 
part $(h\cdot \alpha^*)\,\alpha$,
\begin{equation}
y=\frac{h\cdot H }{h\cdot\alpha }-\frac{\alpha\cdot H}{\alpha^2}.
\end{equation}
Clearly, the 4 bosonic and 2 fermionic adjoint zero modes around such an 
embedded $\alpha^*$ monopole should arise from the $SU(2)_\alpha$ triplet
sitting at the weights $\alpha,0,-\alpha$. But for them, both the mass
term $h'\cdot H$ and the hypercharge $y$ vanish. These zero-modes are
obtained from the $SU(2)$ zero-modes by a straightforward embedding.
Obviously the same is true for zero modes around any embedded monopole
of higher magnetic charge $k\alpha^*$.

There could be further fermionic zero modes from hyper multiplets.
Consider $\alpha=\beta_{ij}^+=(e_i+e_j)/\sqrt 2$. The elementary 
fields fall into various representations with respect to $SU(2)_{\beta_{ij}^+} 
$. In particular, there are a pair of Dirac fields in the  $SU(2)_{
\beta_{ij}^+}$ triplet, sitting at the three weights $\beta_{ij}^+,
0,-\beta_{ij}^+$. One Dirac field belongs to the adjoint
vector multiplet while the other belongs to the antisymmetric hyper multiplet.
Again by an $SU(2)$ embedding, it is clear that the former contains the two
adjoint fermionic zero modes that are supersymmetric partners of the bosonic
zero modes. At the same time, the latter also must contain two fermionic zero 
modes by virtue of having the identical coupling to the $(\beta_{ij}^+)^*$ 
monopole. Furthermore, since the weights of the antisymmetric multiplet
are a subset of the adjoint multiplet, there cannot be any other zero modes
from this hyper multiplet. Finally, there are 8 $SU(2)_{\beta_{ij}^+}$ 
doublet Dirac fields from the 4 fundamental hyper multiplet, sitting at
$\gamma_i/2,-\gamma_j/2$ or at $-\gamma_i/2,\gamma_j/2$. No fermionic
zero mode results from them, however. If there were such an zero mode,
there must also be zero modes from a triplet sitting at $\gamma_i,
\beta_{ij}^-, -\gamma_j$ or at $-\gamma_i,-\beta_{ij}^-,\gamma_j$. {}From 
the adjoint zero mode counting, we know there is no such zero mode.

Thus we learn that the $(\beta_{ij}^+)^*$ monopole carries four fermionic 
zero modes in all, two from the adjoint vector multiplet, two more from 
the antisymmetric hyper multiplet and none from the fundamental 
hyper multiplets. Exactly the same reasoning
goes for $\alpha=\beta_{ij}^-=(e_i-e_j)/\sqrt 2$, so the $(\beta_{ij}^-)^*$
monopole also carries four such zero modes.

The $\gamma_i^*$ monopole exhibits a somewhat different behavior. 
Again, the adjoint vector multiplet contributes a $SU(2)_{\gamma_i}$
triplet Dirac field, sitting at $\gamma_i,0,-\gamma_i$, which in turn 
contains all the necessary fermionic partners of the bosonic zero modes. 
Similarly as above, this implies that there are no fermionic zero modes
associated with any other adjoint weights including $\beta_{ij}^\pm$,
and from this we deduce that the antisymmetric hyper multiplet contributes
no zero modes in the $\gamma_i^*$ monopole background. Finally, 
the four fundamental hyper multiplets decompose into four  $SU(2)_{\gamma_i}$
doublets, sitting at $\gamma_i/2,-\gamma_i/2$ 
and remaining $8(n-1)$ singlets. Each doublet gives rise to 
a single fermionic zero mode. Thus, the $\gamma_i^*$ monopole carries 
six fermionic zero modes in all, two from the adjoint vector multiplet,
none from the antisymmetric hyper multiplets, and four from the four 
fundamental hyper multiplets. Note that for these modes also, the mass term
$h'\cdot H$ and the hypercharge $y$ vanish.

Finally note that, for a unit-charged $\alpha^*$ monopole, 
the fermionic zero modes from an $SU(2)_\alpha$ triplet
carry spin $1/2$ while those from the doublet are spinless. This is
due to the well-known shift of angular momentum in the electric-magnetic
bound system. To summarize, the monopole carries four bosonic zero modes 
in a generic vacuum, and in addition carries the fermionic zero modes 
as described in Table 1. It is worthwhile to repeat the fact that all of
this zero modes are obtained from a simple $SU(2)$  embedding. 

\hskip 2.5cm
\begin{tabular}{|l|l|l|}\hline
monopole charge    & matter multiplet    & \# of zero modes (spin) \\ \hline
                   & adjoint vector      & 2 ($1/2$)   \\ \cline{2-3}
$(\beta_{ij}^\pm)^*=\beta_{ij}^\pm$ 
                   & antisymmetric hyper & 2 ($1/2$)   \\ \cline{2-3}
                   & fundamental hypers   & 0             \\ \hline
                   & adjoint vector      & 2 ($1/2$)   \\ \cline{2-3}
$\gamma_i^*=\gamma_i/2$       
                   & antisymmetric hyper & 0             \\ \cline{2-3}
                   & fundamental hypers  & 4 (0)         \\ \hline
\end{tabular}

\begin{quote}
{\bf Table 1:} The number of fermionic zero modes from various matter
fields. Because the two Higgs are completely misaligned, each monopole
carries four bosonic zero modes.
\end{quote}

\subsection{Dyonic Spectrum}

We have proposed in the introduction that our theory possesses an exact
$SO(8)\sdtimes SL(2,Z)$ duality. If this is indeed the case, the BPS spectrum
should have infinite towers of dyons built on every  elementary particle
listed in Eq.~(\ref{particle}). In the case of the completely 
misaligned Higgs expectation value, we will see now that the
semiclassical spectrum can be obtained easily from the known results from 
the $N=4$ and $N=2$ cases with $SU(2)$ gauge symmetry. In this case,
the S-duality passes the test quite easily.

For  $(\beta_{ij}^{\pm})^*=\beta_{ij}^{\pm}$, we expect the electric
vector meson and the magnetic soliton, or rather their respective 
supermultiplets, are dual to each other.
The spectrum of elementary electric sector of charge $\beta_{ij}^{\pm}$ is 
made of a vector  and a hyper multiplets, whose state content is identical to
a $N=4$ supermultiplet. Furthermore, since they belong to either the adjoint
or the antisymmetric tensor, these multiplets are invariant under the 
flavor symmetry $SO(8)$.  

In the previous section we showed that a single $(\beta_{ij}^\pm)^*$ monopole 
has four bosonic zero modes and four complex fermionic zero modes (forming 
two spin doublets).
Although the two of the four fermionic zero modes are from the adjoint
Dirac  fields and the other two from the antisymmetric tensor, they solve
the same $SU(2)_{\beta_{ij}^{\pm}}$ triplet zero mode equation. In effect, it
is as if these fermionic zero modes arise from the two adjoint Dirac fields
of $N=4$ $SU(2)$ theory. The upshot is then, the monopole dynamics of $k$
identical $(\beta_{ij}^{\pm})^*$ monopoles of this $N=2$ system is identical
to that of $k$ monopoles in  $N=4$ $SU(2)$ system. The latter is known to
have a $SL(2,Z)$-invariant tower of $(n_m=p,n_e=q)$ dyons with all co-prime 
integer pairs \cite{sen,segal}. It follows that there is a similar 
$SL(2,Z)$-invariant dyonic tower of charges $(p\beta_{ij}^\pm,
q\beta_{ij}^\pm)$.

Furthermore, the supermultiplet corresponding to each dyon must be 
effectively a $N=4$ vector multiplet, which in turn consists of a $N=2$
vector  and a hyper multiplet. Thus, on each weight $\beta_{ij}^{\pm}$,
two $(p,q)$ dyonic towers sit, one in the vector multiplet and the other
in the hyper multiplet. This is exactly the BPS spectra one obtains by
acting $SL(2,Z)$ on the elementary spectrum in 
Eq.~(\ref{particle}).

As $\gamma_i^*=\gamma_i/2$, a single monopole of
magnetic charge $\gamma_i^*$ should be a dual to quarks of electric
charge $\gamma_i/2$. As shown in Table 1, there are four bosonic zero
modes, and two fermionic zero modes from the adjoint Dirac spinor, and
four fermionic zero modes from the fundamental Dirac spinor.  This
system is exactly like the monopole system that arises in the $N=2$
scale-invariant $SU(2)$ model of Seiberg and Witten \cite{seiberg}. 
Similarly as above, we can pretend 
that we are considering the multi-monopole dynamics of the 
scale-invariant Seiberg-Witten model, 
and the S-duality of $\gamma_i^*$ dyons simply should follow from the 
S-duality of the BPS spectrum of the latter model. 

The monopole dynamics of the $N=2$ model is a bit more involved than
the $N=4$ case, and so far one and two monopole dynamics were
considered in detail.  Dyons with unit magnetic charge
$\gamma_i^*=\gamma_i/2$ must come in four hyper multiplet, or eight
half-hyper multiplets, transforming under $SO(8)$ as either $8_c$ or
$8_s$. The two fermionic zero modes from the adjoint vector simply
fill out a half-hyper multiplet, so one should find the degeneracy
factor $8$ from the fundamental vector bundle.  One would have thought
that the $4$ fundamental zero modes would generate $2^4=16$
degeneracy. The reason why one obtains either $8_c$ or $8_s$, can be
found in the nontrivial $O(1)$ bundle \cite{manton,seiberg}.  Because
the fundamental zero modes acquire a phase of $-1$ upon a $Z_2$ gauge
transformation, the $\gamma_i^*$ electric charge can be either even or
odd depending on whether even or odd number of the fundamental zero
modes are excited. Or equivalently, even (odd) charged dyon can have
only even (odd) number of fermions excited.  In this way, it was found
\cite{seiberg,gaunt1} that the dyons of charge
$(\gamma_i^*,2q\gamma_i^*)$ belong to $8_s$ of the $SO(8)$ flavor
symmetry and those of $(\gamma_i^*,(2q+1)\gamma_i^*)$ belong to $8_c$.
They are related to the quark states of charge $(0,\gamma_i/2)=(0,
\gamma_i^*)$ in $8_v$ by $SL(2,Z)$ transformations. As was stated
above, the duality group is a semidirect product $SO(8)\sdtimes
SL(2,Z)$ where some $SL(2,Z)$ generators act as a triality.
 
Two identical monopoles of magnetic charge $\gamma_i^*$ would have the
magnetic charge $\gamma_i$. The fermionic zero modes of this magnetic
sector would be doubled from those for a single monopole. Their dyonic
spectrum is again exactly as in the scale-invariant Seiberg-Witten
$SU(2)$ theory. By establishing an index theorem in the $O(2)$ index
bundle~\cite{manton} of the fermion zero modes from the fundamental
fermions on the Atiyah-Hitchin space~\cite{atiyah}, the dyonic
spectrum of this sector has been shown to be realized as bound states
of right quantum numbers \cite{gaunt1}.  The dyons of charge
$(2\gamma_i^*,(2q+1)\gamma_i^*)$ are in 8 half-hyper multiplets,
transforming as $8_v$ of $SO(8)$, and should be considered a part of
the dyonic tower sitting at $\gamma_i^*$.  The dyons of charge
$(2\gamma_i^*, 2q\gamma_i^*)=(\gamma_i,q\gamma_i)$, on the other hand,
come in the vector multiplet and invariant under $SO(8)$. They are
the dual images of the gauge vector multiplet of electric charge
$\gamma_i$.

\section{Monopole Dynamics with Aligned Higgs}

The question of monopole dynamics and the BPS spectrum become far more
subtle when the two Higgs expectation values are aligned within the
Cartan subalgebra.  This is because of the additional bosonic and
fermionic zero modes that emerge in such cases. With two Higgs
expectation $h$ and $h'$, we have seen that
\begin{equation}
M_{\mu^*+\nu^*}=M_{\mu^*}+M_{\nu^*} ,
\end{equation}
if and only if $(h\cdot\mu^*)(h'\cdot\nu^*)=
(h'\cdot\mu^*)(h\cdot\nu^*)$.  Thus if the two Higgs are aligned in
the ($\mu,\nu$) plane, a monopole of charge $\mu^*+\nu^*$ can be split
into a pair of monopoles of charges $\mu^*$ and $\nu^*$ without
costing any energy. This translates into additional massless degrees
of freedom in the low energy monopole dynamics, and the moduli space
dynamics becomes more involved.

While one may expect that the S-duality of the BPS spectrum in generic
vacuum will continue to hold in such exceptional cases, the $N=2$
supersymmetric theories are known to exhibit discontinuity in
spectra as one moves around the vacuum moduli space. An explicit check
is necessary.  If the S-duality holds everywhere, this would imply
that certain sets of monopoles must combine quantum mechanically to
give rise to supersymmetric quantum bound states of specific
nature. While we could not construct all such bound states necessary,
there turned out to be many cases where the existence of the bound
states and even their precise forms can be inferred from the known
results in $N=4$ Yang-Mills theory.  We start the section by
revisiting the matter of zero modes, and attack the problems of
various dyon spectra, weight by weight. For notational simplicity, we
will consider the case where $h$ is parallel to $h'$ in the entire
Cartan subalgebra.

\subsection{Zero Modes Revisited}

When the two Higgs expectations are aligned, one can rotate $h+ih'$ by 
the global $U(1)$ $R$-symmetry so that we may effectively set 
one of the Higgs expectation, say $h'$, to zero and subsequently let the
corresponding Higgs field $\Phi'$ vanish as well. 
The spinorial zero-mode equation of Section 3.1 is then reduced to
\begin{equation}
[\gamma^i D^i+\gamma^5 \Phi_\alpha]\, 
\Psi+\gamma^5(h\cdot\alpha)\,y\,\Psi=0.
\end{equation}
Recall that the hypercharge $y$  reflects the discrepancy between
the Higgs expectation $h$ and its $SU(2)_\alpha$ part $(h\cdot
\alpha^*)\,\alpha$.  The number of zero modes is determined by the
$SU(2)_\alpha$ representation and the hypercharge, and  Table 2
summarizes the results obtained by Weinberg \cite{weinberg}.

\hskip 3cm
\begin{tabular}{|l|r|r|}\hline
$SU(2)_\alpha$ representation & hypercharge & \# of zero modes \\ \hline
singlet     &      any $y$    & 0            \\ \hline
doublet     &      $1/2>|y|$  & 1            \\ \cline{2-3}
            &      $|y|\ge 1/2$ & 0          \\ \hline
triplet     &      $1> |y| $  & 2           \\ \cline{2-3}
            &      $|y|\ge 1$ & 0          \\ \hline
\end{tabular}

\begin{quote}
{\bf Table 2:} The number of Dirac zero modes for an $SU(2)$ embedded,
spherically symmetric monopole of charge $\alpha^*$. The $SU(2)_\alpha$ 
representation is that of the Dirac spinor.
\end{quote}

Now suppose we arranged the Cartan generators such that the Higgs expectation
$h$ satisfies the following inequality,
\begin{equation}
0<h_1<h_{2}<\cdots <h_{n-1}<h_n.
\end{equation}
Then a set of $n$ roots $\{\mu_i\}=\{\gamma_1,\beta^-_{21},\dots,
\beta^-_{n,n-1}\}$ has a natural interpretation as the positive
simple roots. Consequently,
a $\mu_i^*$ monopole has exactly 4 bosonic zero modes so that one can
interpret it as a fundamental soliton. For monopoles of total charge
$k\mu_i^*$ with any positive integer $k$, the zero mode counting
remains unchanged from the previous section. Others will acquire more 
bosonic and fermionic zero modes, and we may proceed with the counting 
with the  help of Table 2,
keeping in mind that the number of the zero modes is additive in this
case of completely aligned Higgs.
Table 3 summarizes the number of fermionic zero modes for monopoles of 
various charges, assuming that $h$ is parallel to $h'$ in the unitary gauge. 
Some of the zero modes will be lifted if $h$ is partially misaligned with
$h'$.

\hskip 1.5cm
\begin{tabular}{|l|r|r|r|}\hline
monopole charge    & matter multiplet    & \# of zero modes  & bundle structure
\\ \hline
               & adjoint      & $(2i-2j)k$   & $Sp(2ik-2jk-2)$ \\ \cline{2-4} 
$k(\beta_{ij}^-)^*=k\beta_{ij}^-$ 
               & antisymmetric  & $(2i-2j)k$   & $Sp(2ik-2jk-2)$\\ \cline{2-4}
               & fundamental    & 0        &  $\cdot$  \\ \hline
               & adjoint        & $2ik$      & $Sp(2ik-2)$ \\ \cline{2-4}
$k\gamma_i^*=k\gamma_i/2$      
               & antisymmetric  & $(2i-2)k$    & $Sp(2ik-2k)$ \\ \cline{2-4}
               & fundamental    & $4\times k$ & $O(k)$        \\ \hline
               & adjoint        & $(2i+2j)k$   & $Sp(2ik+2jk-2)$\\ \cline{2-4}
$k(\beta_{ij}^+)^*=k\beta_{ij}^+$ 
               & antisymmetric  & $(2i+2j-4)k$ & $Sp(2ik+2jk-4k)$\\ \cline{2-4}
               & fundamental    & $4\times 2k$ & $O(2k)$          \\ \hline
\end{tabular}
 
\begin{quote}
{\bf Table 3:} The number of fermionic zero modes from various matter
sectors. The number of bosonic zero modes is twice that of the fermionic
adjoint zero modes. The last column shows the structure group of the vector
bundle spanned by these zero modes. See Section 4.4 for detailed explanations.
\end{quote}

\subsection{The $\mu_i^*$ Dyons}

As observed above, the $n$ roots $\{\mu_i\}=\{\gamma_1,\beta^-_{21},\dots,
\beta^-_{n,n-1}\}$ correspond to the $n$ fundamental monopoles. The simplest
dyon spectra one can address are those
of magnetic charges $k\mu_i^*$. But for these fundamental dyons,
the zero mode counting and the moduli space dynamics are identical to 
that of the previous section, and we need not ask anything new.
The S-duality of $\mu_i^*$ dyons (and 
$2\gamma_1^*$ dyons as well) in the aligned Higgs case follows from that
of the misaligned Higgs case.

\subsection{The $(\beta^-_{ij})^*$ Dyons}

Note the following very useful fact about the $(\beta^-_{ij})^*=
\beta^-_{ij}$ monopoles. First, the roots $\{\beta_{ij}^-,
-\beta_{ij}^-\}$, together with the $(H_i-H_{i+1})$'s, span an $SU(n)$
subgroup of the $Sp(2n)$. Any BPS monopole of charge
$(\beta_{ij}^-)^*$ can be subsequently regarded as an embedding of a
$SU(n)$ monopole of the same charge to the larger $Sp(2n)$ gauge
group, again up to a uniform part of the Higgs field proportional to
$\sum_i H_i$. Then, the adjoint representation of $Sp(2n)$ is
decomposed into an adjoint, a symmetric tensor and its conjugate, as
well as a singlet of the said $SU(n)$ subgroup, while the
antisymmetric tensor of $Sp(2n)$ consists of an adjoint, an
antisymmetric tensor and its conjugate of the same $SU(n)$.  But in
fact, we know from the zero-mode counting of the $SU(n)$ monopole that
all $2i-2j$ zero modes arise from the adjoint of the $SU(n)$ alone, as
a $(\beta_{ij}^-)^*$ monopole is made of $i-j$ distinct fundamental
monopoles. Consequently, all the antisymmetric
zero modes must fall into the adjoint of this $SU(n)$ as well.

Since the $\sum_iH_i$ part of the Cartan algebra commutes with the
embedded $SU(n)$, this implies that the fermionic zero modes from the
vector multiplet solve exactly the same equations as those from the
antisymmetric hyper multiplet do.  Effectively, the number of the
fermionic zero modes from the vector multiplet is doubled. But we are
already familiar with such a multi-monopole dynamics. It is a
multi-monopole dynamics obtained from $N=4$ $SU(n)$ Yang-Mills theory
(with all six Higgs aligned), which has been studied thoroughly by the
authors as well as others \cite{gaunt2,gibbons}.  By now it is widely
believed and comprehensively tested that such an $N=4$ theory
possesses fully S-dual BPS spectrum at the semiclassical level.  In
other words, the well-known S-duality of $N=4$ $SU(n)$ Yang-Mills
theory automatically implies that there is an $SL(2,Z)$-invariant
dyonic tower of charges $(p\beta^-_{ij},q\beta^-_{ij})$ for all
co-prime integer pair $(p,q)$. Explicit forms of the bound states is
known for $p=1$ \cite{gaunt2,gibbons}.

Finally, because the monopole dynamics is effectively that of the
$N=4$ Yang-Mills theory, the bound states one gets are in the $N=4$
vector multiplet. An $N=4$ vector multiplet is in turn composed of one
$N=2$ vector multiplet and one $N=2$ hyper multiplet, so we know that
on each weight $\pm (\beta^-_{ij})^*= \pm \beta^-_{ij}$ there are a
pair of $(p,q)$ dyonic towers, one in vector and the other in hyper.
This is exactly the spectrum one would obtain by performing $SL(2,Z)$
transformation on the elementary BPS spectra on $\pm \beta^-_{ij}$
delineated in Eq.~(\ref{particle}).

\subsection{The Index Bundle}

Before proceeding further, we need to elaborate on some basic facts
about the moduli space dynamics. The zero modes can be thought of
deviations from the classical monopole background. For bosonic zero
modes from the vector multiplet, this is especially clear since they
literally encode small deformation of the classical BPS monopole
solution. The fermionic zero modes deform the state at quantum level,
on the other hand, by supplying an additional set of fermionic
harmonic oscillators to the moduli space quantum mechanics.  One
consequence is that the wavefunction becomes a multi-component variety,
each component being characterized by which subset of these
oscillators is  excited.

The extra interactions introduced through such fermionic excitations 
arise naturally from the field theory as follows. Consider a 
Dirac Lagrangian,
\begin{equation}
{\cal L}=\int d^4x\;\left\{i\bar\Psi\gamma^\mu D_\mu \Psi+\cdots\right\}.
\end{equation}
When a background Yang-Mills field carries $m$ number of $\Psi$ zero modes,
say $\Psi_{p}$, $p=1,...,m$, their contribution to the moduli space dynamics
is found by expanding $\Psi=\sum \eta_p(t)\Psi_{p}$ with the time-dependent,
complex Grassmannian, collective coordinates $\eta_p$, and inserting it back 
to the Lagrangian. Only the time derivative piece survives, and one finds
\cite{manton,blum}
\begin{equation}
{\cal L} \quad \rightarrow \quad\int dt\;\left\{i\tilde\eta_p
\frac{d}{dt}\eta_p+i\tilde\eta_p \eta_q {\cal A}_{apq}\frac{dz_a}{dt}+
\cdots\right\},
\end{equation}
where the tilde denotes the complex conjugation. The 
$z_a$'s are the bosonic collective coordinates that parameterize the
background Yang-Mills field, and the ``connection'' ${\cal A}$ is given by
\begin{equation}
{\cal A}_{apq}=\int dx^3\; \Psi_{p}^\dagger
\delta_a\Psi_{q}.
\end{equation}
This connection is easily seen to be anti-hermitian, ${\cal A}_{apq}=-
\tilde{\cal A}_{aqp}$. In this sense, the $m$ fermionic zero modes
can be regarded as spanning a $U(m)$ vector bundle on the moduli space. 

When the gauge group representation of $\Psi$ is real or pseudo-real, however,
there is a further constraint on ${\cal A}_{apq}$. The Dirac operator is then
equivalent to its charge conjugate up to a redefinition of the spinor,
\begin{equation}
\Psi\quad\Rightarrow\quad \Psi_c = S\otimes V\, \tilde\Psi .
\end{equation}
The unitary matrix $S$ acts on the group indices and is either symmetric if 
the group  representation is real or antisymmetric  if the group 
representation is pseudo-real. Alternatively, $S\tilde S=\pm 1$ for real and
pseudo-real representations, respectively. The 
unitary matrix $V=-V^T$ is the charge conjugation matrix for the $SO(3,1)$
Dirac spinor. This means in particular that if $\Psi_p$ is a zero mode, 
so is $(\Psi_p)_c$. Thus, the two must be related by yet another unitary
transformation ${\cal C}$,
\begin{equation}
(\Psi_p)_c= {\cal C}_{pq}\Psi_q,
\end{equation}
where ${\cal C}\tilde{\cal C} =\mp 1$, or equivalently ${\cal C}=\mp
{\cal C}^T$, for $S\tilde S=\pm 1$. A unitary redefinition of the
basis $\Psi_p\Rightarrow \Psi_q\tilde U_{qp}$ rotates ${\cal C}$ to
$U^T {\cal C} U$, which clearly preserves ${\cal C}\tilde{\cal
C}$. Finally, taking the determinant of ${\cal C} \tilde{\cal C}$, we
learn that the number of zero modes must be even, $m=2k$, whenever
${\cal C}\tilde{\cal C}=- 1$.

Using this charge conjugation twice on ${\cal A}_a$, we find the following 
reality constraint,
\begin{equation}
{\cal A}_a{\cal C}+{\cal CA}_a^T=0.
\end{equation}
The unitarity guarantees that $\cal C$ can be brought into a
(skew-)diagonalized form by an orthogonal $U$, and then ${\cal C}\tilde{\cal
C}=\mp 1$ tells us that the (skew-)eigenvalues are pure phases. These 
phases are easily absorbed by the zero modes themselves. The 
canonical form of $\cal C$ is then
\begin{eqnarray}
{\cal C}&=&1\otimes i\tau_2 \quad \hbox{if the group representation 
is real}, \\
{\cal C}&=& 1\quad \hskip 9mm \hbox{if the group representation is 
pseudo-real}.
\end{eqnarray}
We can easily recognize these as the invariant bilinear forms of
$Sp(m=2k)$ and $O(m)$, respectively. Thus, when the fermions are in real and
pseudo-real representations of the gauge group, the structure group of
the index bundle reduces to $Sp(m=2k)$ and $O(m)$, respectively.

The adjoint representation of any gauge group is real, so we should find 
a $Sp(2k)$ bundle to emerge from the $2k$ fermionic zero modes of the 
vector multiplet. Because of the $N=2$ supersymmetry, on the other hand, 
this bundle is equivalent to the co-tangent bundle of the $4k$-dimensional 
monopole moduli space, so 
the latter must have the structure group $Sp(2k)$. A $4k$-dimensional
manifold with a structure group $Sp(2k)$ is by definition hyperk\"ahler, 
and thus we have rediscovered the well-known fact that the moduli space is 
naturally hyperk\"ahler. In fact, the actual structure group of the 
moduli space of $4k$ dimensions is somewhat smaller, $Sp(2k-2)$, because
the 4-dimensional, flat center-of-mass part factors out in the metric
locally. Thus, by supersymmetry, $2k$ fermionic zero modes of the 
vector multiplet is decomposed as $2+(2k-2)$, where the two associated
with the center-of-mass coordinates span a flat bundle, while the
remaining $(2k-2)$ span a $Sp(2k-2)$ vector bundle over the relative
moduli space.

The fundamental representation of $Sp(2n)$ is pseudo-real, which tells us
that the resulting index bundle is a $O(m)$ vector bundle. This generalizes
the observation by Manton and Schroers that the zero mode bundle from a
fundamental representation of $SU(2)=Sp(2)$ is a real vector bundle 
\cite{manton}. This
fact was a crucial ingredient in studying the S-dual spectrum of 
scale-invariant $N=2$ $SU(2)$ model.

The most interesting aspect of the Yang-Mills theory we are considering in
this note, is the presence of the antisymmetric matter. The group 
representation is also real, so $2k'$  fermionic zero modes
span a $Sp(2k')$ vector bundle in general. Note that the case of
$(\beta^-_{ij})^*$ monopoles in Section 4.3 is an exceptional case;
instead of spanning $Sp(2i-2j)$ as the structure group, as the general
argument would imply, the $2k'=2i-2j$, zero modes turned out to span a
$Sp(2i-2j-2)$ bundle. This happened because the antisymmetric zero modes 
are actually identical to the adjoint ones. Recall that the structure group
of the adjoint index bundle with $2k=2i-2j$ zero modes is $Sp(2i-2j-2)$ 
because of the moduli space decomposition.

\subsection{The $\gamma_2^*$ Dyons}

The next nontrivial problem is that of $\gamma_2^*= \gamma_2/2$ dyons. 
The purely magnetic state must be realized as a threshold bound 
state of a $\gamma_1^*$ monopole and a $(\beta^-_{21})^*$ monopole.
Is the monopole dynamics again effectively that of $N=4$ Yang-Mills as in
the $(\beta^-_{ij})^*$ dyons? Not quite so. The $\gamma_2^*$ monopole 
carries $4$ adjoint zero modes, 2 antisymmetric tensor modes 
and $4\times 1$ fundamental  zero modes. There is no obvious doubling of 
the adjoint fermionic modes. Nevertheless, the bound state problem can be
mapped to an $N=4$ problem as follows.

Although the adjoint multiplet has twice the number of zero mode as the
antisymmetric one, their structure groups are both $Sp(2)$, because  of
the usual decomposition of the moduli space ${\cal M}$ \cite{gaunt2},
\begin{equation}
{\cal M}=R^3\times \frac{R^1\times {\cal M}_0}{ Z}.
\end{equation}
Of the four adjoint modes, two are supersymmetric partners of the 
center-of-mass coordinates parameterizing $R^3\times R^1$. These two
do not participate in the interaction and simply provide the 
$N=2$ BPS supermultiplet structure to the wavefunction.
As far as the mutual monopole interaction goes, we have a pair
of $Sp(2)$ vector bundles on ${\cal M}_0$, one from the adjoint vector
multiplet and the other from the antisymmetric hyper multiplet.
For a pair of distinct fundamental monopoles, the relative moduli space
${\cal M}_0$ is always the Taub-NUT manifold \cite{connell,gaunt2}, 
the metric on which can be written as 
\begin{equation}
{\cal G}=\left(1+\frac{1}{r}\right)\,[dr^2 +r^2\sigma_1^2+r^2\sigma_2^2]
+\frac{1}{1+1/r}\,\sigma_3^2 
\end{equation}
up to an overall constant. The 3 one-forms $\sigma_i/2$ are an orthonormal
basis on a unit three sphere, which is invariant under the spatial
rotation $SU(2)$.
In terms of the Euler angles $\theta,\varphi,\psi$, they are
\begin{eqnarray}
& & \sigma_1= -\sin\psi d\theta +\cos\psi\sin\theta d\varphi, \nonumber \\
& & \sigma_2=\cos\psi d\theta +\sin\psi\sin\theta\varphi, \nonumber \\
& & \sigma_3 =d\psi+\cos\theta d\varphi.
\end{eqnarray}
Because the Taub-NUT manifold is simply 
connected, a nontrivial twisting under a $O(1)$, if any, can occur only 
through the modding by the integer group, and fundamental zero modes decouple
from the relative dynamics also. Thus, only the adjoint and antisymmetric 
zero-modes enter the relative moduli space dynamics. 

Let $(\eta^1,\eta^2)$ be the two complex collective coordinates associated 
with the anti-symmetric zero modes. Also let $\psi^a$ be the four real,
supersymmetric partners of the bosonic  collective coordinates $z^a$.
The relative part of the moduli space dynamics is governed by the 
supersymmetric Lagrangian,
\begin{equation}
\int dt\:\left\{ \frac{1}{2}{\cal G}_{ab}\dot z^a\dot z^b+\frac{i}{2}\psi^a
D_t\psi^a+i\eta^\dagger{\cal D}_t\eta + \frac{i}{2}\eta^\dagger
[{\cal F}_{ab}\psi^a\psi^b]\eta\right\},
\label{lagra}
\end{equation}
where the two covariant derivatives
\begin{eqnarray}
D_t\psi^a &=& \partial_t\psi^a+\dot z^c w_{abc}\psi^b ,\nonumber \\
{\cal D}_t\eta^p &=& \partial_t\eta^p+\dot z^c {\cal A}_{cpq}\eta^q ,
\end{eqnarray}
are defined with respect to the connections of the co-tangent bundle and 
the index bundle of the anti-symmetric zero modes, respectively. ${\cal F}$ 
is the curvature of the latter, $d{\cal A}+{\cal A}^2$. The 
quantization of the $\psi^a$'s dictates that the wavefunction should be
a Dirac spinor on ${\cal M}_0$ \cite{gaunt1}, while $\eta$ quantization 
leads either to one of two $Sp(2)$ singlets or to a $Sp(2)$ doublet. 
More explicitly, the canonical quantization of $\eta$ leads to a pair of 
fermionic harmonic oscillators,
\begin{equation}
\{\eta_i,\tilde\eta_j\}=\delta_{ij}.
\end{equation}
The vacuum $|0\rangle$ is defined to be annihilated by the $\eta_i$'s, then, 
the four states are $|0\rangle$, $\tilde\eta_1\tilde\eta_2 |0\rangle$, 
$\tilde\eta_1|0\rangle$, and $\tilde\eta_2 |0\rangle$. The last two form a
doublet, while the first two are singlets. The Hamiltonian 
is the square of a supersymmetry generator ${\cal Q}$, which is in effect
a Dirac operator on ${\cal M}_0$ \cite{blum,gaunt1}.

In order to find the bound states in the relative moduli space,
we need to know the $Sp(2)$ connection ${\cal A}$ over  ${\cal M}_0$. 
The moduli space dynamics inherits the symmetries of the underlying 
monopole dynamics, namely a rotational $SU(2)$ and a gauge $U(1)$ 
\cite{gaunt1,gaunt2}, and $\cal F$ must respect them as well. Furthermore,
a theorem due to Hitchin says the curvature must be anti-self-dual in
the space of differential forms \cite{manton}, so the $Sp(2)$ connection
$\cal A$ must be that of $SU(2)\times U(1)$-invariant anti-instanton(s) 
on ${\cal M}_0$. It turns out that there is a one-parameter family of such 
$Sp(2)$ connections, up to gauge transformations,
\begin{equation}
{\cal A}_a dz^a=\left[1-\frac{1}{1+r/R}\right] \left\{\sigma_1 T^1+
\sigma_2 T^2\right\} + \left[1-\frac{1}{(1+r/R)(1+r)}\right]\,\sigma_3 
T^3,
\end{equation}
where $[T^j,T^k]=-\epsilon_{ijk}T^i$ are the anti-hermitian $Sp(2)=SU(2)$ 
generators, and the size parameter $R$ is positive. This connection is 
invariant under the $U(1)$ isometry which rotates the $(\sigma_1,\sigma_2)$ 
pair, as long as one rotates the $(T^1, T^2)$ pair simultaneously.

In fact, the connection $w$ of the Taub-NUT manifold ${\cal M}_0$ can
also be regarded as a special case of this one parameter family. This
has to be the case, since $w$ is also effectively an $Sp(2)$
connection whose curvature is anti-self-dual and invariant under the
isometry of ${\cal M}_0$. The self-dual part $w^{(+)}$ is curvature
free, and can be transformed away in one of the two $SU(2)$'s in
$SO(4)=SU(2)\oplus SU(2)$. In contrast, the anti-self-dual part
$w^{(-)}$ takes values in the other $SU(2)$, and must be interpretable
also as an $SU(2)=Sp(2)$ anti-instanton. A direct computation shows
that the $w^{(-)}$ coincides with ${\cal A}_a dz^a$ with $R=1$, after
an appropriate reinterpretation of $T^i$ as Euclidean Lorentz
transformation generators.  Alternatively, starting with an
$Sp(2)=SU(2)$ bundle with the connection $w^{(-)}$, we can build an
$SO(4)$ vector bundle by swapping the two complex coordinates similar
to the $\eta$'s for four real ones. By gauge rotating the curvature
free part of the $SO(4)$, which mixes holomorphic and anti-holomorphic
coordinates, we may induce a curvature free $w^{(+)}$, and once this
is done the result is simply the co-tangent bundle on the Taub-NUT. In
fact, this is one way to recover the adjoint part of the Lagrangian
starting with two fermionic zero modes from the vector multiplet.

The Pontryagin number of $\cal A$ is easily seen to be $-1$, and this
is an anti-instanton of arbitrary ``size'' $R$. Can the anti-instanton
be of an infinite size? Unless $R=\infty$, the asymptotic form of
${\cal A}$ on the surface $S^3$ spanned by the $\sigma_i$'s as
$r\rightarrow \infty$ is pure gauge of unit winding number, and can be
removed by a large gauge transformation on $S^3$. On the other hand,
if $R=\infty$, the anti-instanton becomes infinitely large and $\cal
A$ degenerates to a $U(1)$ connection $r/(1+r)\sigma_3T^3$ whose field
strength ${\cal F}\sim -T^3\sin\theta\, d\theta\, d\phi$ is not a pure
gauge on the asymptotic $S^3$.  A useful thing to note is that, when
$r\rightarrow \infty$, zero modes from the anti-symmetric hyper
multiplet are localized at the $(\beta^-_{21})^*$ monopole and in fact
solve the same equation as the adjoint zero-modes around the
$(\beta^-_{21})^*$ monopole, to the leading order in $1/r$. This,
together with the above identification of $w$ with $\cal A$ at $R=1
<\infty$, implies that the connection of the index bundle of $\eta$
must also be pure gauge on the asymptotic $S^3$. Thus, we learn that
the ``size'' of the anti-instanton must be finite, $R<\infty$, and the
index bundle does not degenerate to a $U(1)$ type.

An independent way to see $R<\infty$ is to consider the spin of the zero
modes. Consider the Noether angular momentum obtained from the
Lagrangian (\ref{lagra}), neglecting the $\psi^a$ modes, by using the
rotation Killing vectors, 
\begin{eqnarray}
& & L_x = -\sin\varphi \frac{\partial}{\partial \theta} +
\frac{\cos\varphi}{\sin\theta} \bigl(
\frac{\partial}{\partial\psi}-\cos\theta
\frac{\partial}{\partial\varphi}\bigr), \nonumber \\ 
& & L_y = \cos\varphi \frac{\partial}{\partial \theta} +
\frac{\sin\varphi}{\sin\theta} \bigl(
\frac{\partial}{\partial\psi}-\cos\theta
\frac{\partial}{\partial\varphi}\bigr), \nonumber\\
& & L_z = \frac{\partial}{\partial\varphi}.
\end{eqnarray}
The anomalous contribution
from $\eta$, in addition to the familiar bosonic part ${\bf J}_B={\bf r}\times 
\dot{\bf r}+q_B\hat{\bf r}$ with $q_B=(\sigma_3/dt)/(1+1/r)$, is given by
\begin{equation}
\Delta J_i= q_\eta \hat r_i+ S_i
\end{equation}
where $q_\eta=-i\tilde{\eta}T^3\eta/[(1+r/R)(1+r)]$  is an $U(1)$
electric charge from $\eta$ and, at large $r$  
with any finite fixed $R$, $S_i$ converges to
\begin{equation}
S_i^\infty=-i\tilde{\eta} U^\dagger T^i U \eta,
\end{equation}
with a large gauge transformation $U$ on $S^3$, which removes the
asymptotic value of the connection ${\cal A}$.  Upon the quantization of 
$\eta$, we find $[S_i^\infty,S_j^\infty]= i \epsilon_{ijk} S_k^\infty$.
Of the four states generated by $\eta$ quantization, the $Sp(2)$ doublet
$\tilde\eta_i|0\rangle$ contributes a quantized spin $1/2$, while the
singlets $|0\rangle$ and $\tilde\eta_1\tilde\eta_2|0\rangle$ contribute
no spin. In such a limit of infinite separation, the two anti-symmetric zero 
modes arise from a triplet of $SU(2)_{\beta_{21}^-}$ and form a spin doublet,
so this behavior of angular momentum is exactly what one should have expected.
On the other hand, if $R=\infty$, the anomalous contribution would be 
of the form,
\begin{equation}
\Delta J_i= q_\eta' \hat r_i
\end{equation}
for a different $U(1)$ charge $q_\eta'$ from $\eta$, and devoid of such a 
spin contribution. (The difference $q_\eta -q_\eta'$ is independently
conserved when $R=\infty$.) This provides a second evidence that $R$ should 
be finite.

Now we are ready to find the threshold bound state of purely magnetic charge
$\gamma_2^*$. Suppose for the sake of argument that $R=1$ and
${\cal A}=w^{(-)}$. The index bundle from the antisymmetric 
hyper multiplet is then equivalent to the co-tangent bundle on the Taub-NUT,
so effectively we have doubled the adjoint fermionic zero modes on the
relative moduli space. The four real degrees of freedom in $\eta$ would
behave exactly as the $\psi^a$'s, after an appropriate redefinition of
the coordinates, and the dynamics is again that of a pair of distinct
fundamental monopoles from $N=4$ Yang-Mills theory. For instance, the 
curvature term in the effective Lagrangian will turn into that of $N=4$
system involving the Riemann curvature of ${\cal M}_0$. Such an $N=4$ system
has been studied previously and a unique bound state is known to arise 
\cite{gaunt2}.\footnote{
In the $N=4$ $Sp(2n)$ theory, a vector multiplet would be generated from
this bound state in ${\cal M}_0$ and becomes dual to certain elementary 
gauge-particles in the GNO-dual \cite{GNO} $SO(2n+1)$ theory. However, the 
center-of-mass part of the index bundles we have differs from the
one found in $N=4$ case, and because of this, the bound state generates
eight half-hyper multiplets, dual to quarks, instead. This explains why 
the gauge group is not mapped to its GNO-dual group but rather to itself.}

What happens if $R\neq 1$? Recall that the bound state in question
must be and is annihilated by the supersymmetry generator $\cal Q$, and thus
contributes to the Witten index.  So the Witten index with an $L^2$
condition is  equal to one. But an index is a topological quantity that
cannot change with small perturbation.  As long as $R$ remains
positive and finite, nothing drastic happens, and the Witten index
remains 1; there must be at least one bosonic bound state.

Furthermore, a careful look at the supersymmetry generator, or equivalently 
a Dirac operator on the Taub-NUT, reveals that in fact there is no extra 
bound state. The wavefunction is a Dirac spinor on the Taub-NUT space
and could be either a singlet or a doublet under the $Sp(2)$. Using 
the Weitzenbock formula, the square of the Dirac operator (which is
the Hamiltonian) can be written as
\begin{equation}
{\cal Q}^2=-\nabla\nabla+\frac{\kappa}{4}+i{\cal F}_{ab}\Gamma^{ab},
\end{equation}
where $\kappa$ is the scalar curvature of Taub-NUT, and $2\Gamma^{ab}=
[\Gamma^a,\Gamma^b]$ with the $SO(4)$ Dirac matrices $\Gamma^a$.
The hyperk\"ahler property ensures $\kappa=0$, while ${\cal F}_{ab}
\Gamma^{ab}$ is null except on anti-chiral part of the spinor in a
$Sp(2)$ doublet. So a nontrivial solution to the Dirac equation can
arise only from anti-chiral spinor in $Sp(2)$ doublet. On the other hand, 
the Witten index in question counts the number of bound states in the form of 
anti-chiral spinor in a $Sp(2)$ doublet or chiral singlet, minus the  
number of those in the form of a chiral $Sp(2)$ doublet or an anti-chiral
singlet. Since only one of the four sectors can contribute to the Witten index,
the index actually counts the total number of the normalizable ground states.

Finally, the supersymmetric wavefunction found at $R=1$ is invariant under 
the $U(1)$ gauge isometry of the Taub-NUT \cite{gaunt2}
and because of this, has  no relative 
electric charge. The electric charge is quantized and cannot change under a 
continuous shift of a parameter, so whatever the actual value of $R$ is,
the bound state in ${\cal M}_0$ is purely magnetic of charge $\gamma_2^*$.
Once we know this, the rest of the quantization may proceed just as in 
the $SU(2)$ case \cite{gaunt1}.\footnote{While decomposition of
the monopole moduli space differs from the $N=2$ $SU(2)$ case in 
Ref.~\cite{gaunt1}, this does not affect the results because the bound 
state in ${\cal M}_0$ is invariant under the discrete modding by 
$Z$.}

One important difference from the $(\beta^-_{ij})^*$ dyon case is the
fermionic zero modes associated with the center-of-mass part 
of the moduli space. Instead of two identical copies of a pair of adjoint 
zero modes from the vector multiplet, as would be appropriate for the actual
$N=4$ system, we have one such pair and four copies of a $O(1)$ bundle
from the four fundamental hyper multiplets. The four copies of $O(1)$ bundle
play the same role as in the $SU(2)$ case, and 
generates $8_s+8_c$ towers of $(\gamma_2^*,
q\gamma_2^*)$ dyons where $8_s$ has even $q$ and $8_c$ has odd $q$.
The remaining two adjoint zero modes associated with the center-of-mass
part carry spin $\pm 1/2$, and generates a short supermultiplet structure
of degeneracy 4, a half-hyper multiplet. For each electric charge
$q$, we conclude, there are $1/2\times 8=4 $ hyper multiplets of charge
$(\gamma_2^*,q\gamma_2^*)$, which 
clearly constitute part of the 4 $SL(2,Z)$-invariant $(p,q)$ dyonic 
towers in hyper multiplet that should be sitting at $\gamma_2^*=\gamma_2/2$.

\subsection{The $\gamma_i^* $ Dyons for $i\ge 3$}

A magnetic charge of $\gamma_i^*$ is a sum of the following fundamental
charges,
\begin{equation}
\gamma_i^*=\gamma_1^*+(\beta^-_{21})^*+(\beta^-_{32})^*+\cdots
+(\beta^-_{i,i-1})^*=\mu_1^*+\mu_2^*+\cdots+\mu_i^* .
\end{equation}
The appropriate multi-monopole moduli space is known \cite{klee1}, so
it only remains to see what the hyper-multiplet zero modes do. There
are still 4 fundamental modes forming four copies of a $O(1)$ bundle,
but thanks to the trivial topology of the relative moduli space, these
do not enter the relative dynamics. The antisymmetric zero modes and
the relative part of the adjoint zero modes each span a $Sp(2i-2)$
vector bundle, and the curvatures of the two bundles enjoy the same
symmetry properties. If the two bundles are actually equivalent, the
relative moduli space dynamics again reduces to $N=4$ dynamics and the
S-duality of $N=4$ Yang-Mills system implies that we recover exactly
one bound state \cite{gibbons}.  Another possibility is that one index
bundle are smooth nonsingular deformations of the other, in which case
we still have the Witten index equal to 1. By quantizing the
center-of-mass part of the moduli space, one would obtain
$(\gamma_i^*, q\gamma_i^*)$ dyons where even $q$'s are in $8_s$ and
each $q$'s are in $8_c$ under $SO(8)$. This is clearly consistent with
the $SL(2,Z)$-invariant $(p,q)$ dyonic towers on $\gamma_i^*$.

Thus, the interesting question is whether one can actually show that the two 
bundles are related by a continuous small deformation, or even equivalent. 
At the moment, we have no compelling argument why that should be so, and 
clearly more study is necessary to address this problem.

\subsection{Others}

So far, we addressed the $(1,q)$ part of the $(p,q)$ dyonic
towers located at $(\beta^-_{ij})^*=\beta^-_{ij}$ and $\gamma_i^*=
\gamma_i/2$. Because of the aligned nature of the Higgs expectations, 
these states appear  already as multi-monopole bound states. Due to the
lack of knowledge on the general multi-monopole dynamics with many
identical monopoles, it appears
that the semiclassical approach is impractical in finding the full 
$(p,q)$ tower of the dyons on these weights.

Also, there are remaining weights, $(\beta^+_{ij})^*=\beta^+_{ij}$
and $2\gamma_i^*=\gamma_i$. In order for the S-duality to hold,
there must be a vector  and a hyper multiplet for each charge $(p\beta_{ij}^+, 
q\beta_{ij}^+)$ for all co-prime integer pairs  $(p,q)$, and similarly 
a $(p\gamma_i,q \gamma_i)$ tower of vector multiplets.
(The existence of the right $2\gamma_1^* =\gamma_1$ dyonic tower 
actually follows from the discussion in Section 3.2, thanks to the
fact $\gamma_1$ is one of the preferred simple root.) The quantization 
of the moduli space dynamics for these is fairly involved. For one thing,
we need certain multi-monopole moduli spaces of so far unknown geometry.
The amount of evidences we collected in the previous two sections, however,
are rather overwhelming, and we have no reason to believe that the  S-duality
should fail for these special cases.

\section{Conclusion}

We have presented several strong evidences for the
$SL(2,Z)\sdtimes SO(8)$ duality in the scale-invariant $N=2$ $Sp(2n)$
theory with one antisymmetric and four fundamental hyper multiplets.
These evidences are found in the semiclassical dyonic spectrum of the 
theory that matches the expected duality-invariant one.
When the Higgs vacuum expectation values are not aligned, the right dyonic 
spectrum for arbitrary $n$ is shown to arise as a consequence of 
well-established S-dualities of the
$N=4$ $SU(2)$ theory and of the scale-invariant $N=2$ $SU(2)$ theory. 
When the two expectation values are parallel, the monopole dynamics are more
involved.
We identified the index bundle on the monopole moduli space, generated 
by the zero modes from the hyper multiplets, and quantized the resulting
moduli space dynamics. Some of the multi-monopole bound states and the
dyonic excitations are constructed, and the result is fully
consistent with the proposed S-duality.

In this note, we ignored the four Higgs from the antisymmetric tensor
hyper multiplet. In the F-theory picture, where the adjoint Higgs encodes
the D3-brane positions in the two compact directions, antisymmetric Higgs 
encode the four position coordinates of the $n$ D3-branes along
the noncompact direction. They will contributes to the masses of dyons
built on $\beta^\pm_{ij}$ weights. But we believe this slight complication 
does not alter the conclusion we drew above.

There are some open questions for the case of the aligned Higgs.
The most obvious ones concern the part
of the spectra which we did not test explicitly, such as $(\beta^+_{ij})^*$ 
dyons, for the lack of knowledge about the moduli space geometry.
On the other hand, there are more accessible problems left unanswered
in this note. For example, we did not determine the precise connection on
the $Sp(2)$ index bundle from the hyper multiplet in Section 4.5 and still 
were able to
deduce the right spectrum. More close study of such nonabelian bundles
on the moduli space should be rewarding in taking the next step and 
uncovering dyons of higher magnetic charges. 

We have started with the F-theory viewpoint of the Yang-Mills system
as the motivation. It gave rise to a particular generalization of the
scale-invariant $SU(2)$ model, and the U-duality of the type IIB
theory in turn strongly suggested that such a generalization would
lead to exactly S-dual $N=2$ theories, which we tested explicitly in
the semiclassical framework. It makes us wonder how far one can go in
realizing $N=2$ and maybe even $N=1$ supersymmetric Yang-Mills systems
as the world-volume theory of D-branes. Recently, there has been a
flurry of research activities in this regard, reproducing known,
sometimes very nontrivial, field theory results or even leading to
completely new insights and systems\cite{witten}. Within the present
context of searching for theories with exact S-duality, it could be
also worthwhile to consider other string vacua and D-branes
configurations and see what conclusions one can draw about resulting
supersymmetric systems.

K.L. is supported by the Presidential Young Investigator Fellowship.
This work is supported in part by U.S. Department of Energy.

\end{document}